\documentclass[12pt,preprint]{aastex}
 \usepackage{graphicx}
\usepackage{natbib}

\shorttitle{Tidally-Induced Apsidal Precession in Double White Dwarf Systems}
\shortauthors{Valsecchi et al.}

\begin{document}
   \title{Tidally-Induced Apsidal Precession in Double White Dwarfs: a new mass measurement tool with LISA}

\author{F. Valsecchi\altaffilmark{1}, W. M. Farr\altaffilmark{1}, B. Willems\altaffilmark{1}, C.J. Deloye\altaffilmark{1}, and V. Kalogera\altaffilmark{1}}
\affil{Center for Interdisciplinary Exploration and Research in Astrophysics (CIERA), and Northwestern University, Department of Physics and Astronomy, 2145 Sheridan Road, Evanston, IL 60208, USA.}


\begin{abstract}

Galactic interacting double white dwarfs (DWD) are guaranteed gravitational wave (GW) sources for the GW detector LISA, 
with more than $10^4$ binaries expected to be detected over the mission's lifetime.
Part of this population is expected to be eccentric, and here we investigate the potential for constraining the white dwarf (WD) properties through apsidal precession in these binaries. We analyze the tidal, rotational, and general relativistic contributions to apsidal precession by using detailed He WD models, where the evolution 
of the star's interior is followed throughout the cooling phase. In agreement with previous studies of zero-temperature WDs, we find that apsidal precession
 in eccentric DWDs can lead to a detectable shift in the emitted GW signal when binaries with cool (old) components are considered. This shift increases significantly for hot (young) WDs. 
We find that apsidal motion in hot (cool) DWDs is dominated by tides at orbital frequencies above $\gtrsim 10^{-4}\,$Hz ($10^{- 3}\,$Hz). 
The analysis of apsidal precession in these sources while ignoring the tidal component would lead to an extreme bias in the mass determination, and could lead us to misidentify WDs as neutron stars or black holes. 
We use the detailed WD models to show that for older, cold WDs, there is a unique relationship that ties the radius and apsidal precession constant to the WD masses, therefore allowing tides to be used as a tool to constrain the source masses.
\end{abstract}


%

\section{Introduction} \label{Intro}
Binary systems containing compact objects constitute an important physical laboratory for investigating the properties of matter under extreme conditions. While direct inference of these properties from ``traditional'' electromagnetic radiation is challenging or impossible, our knowledge about compact object properties can be greatly enhanced by measuring their interactions with their binary companions. The detection of gravitational waves (GWs) from these objects is expected to considerably advance our understanding of astrophysical systems harboring compact objects. Several ground-based GW observatories have already taken data at design sensitivity (e.g., the Laser Interferometer Gravitational-Wave Observatory, LIGO; \citealt{Abbott2007}), while the future generation of GW detectors includes the space-based Laser Interferometer Space Antenna (LISA; Danzmann et al. 1996, \citealt{Hughes2006} and references therein). Our focus here lies on LISA, which was originally designed to be sensitive to GW frequencies between 10$^{-4}$  and  0.1 Hz, and to carry out high-precision astronomy for  sources ranging from sub-solar-mass WDs to million-solar-mass black holes (BH).
In its original design, LISA would have surveyed the whole galactic population of DWDs with period shorter than 6$\,$h, with the potential of detecting from 1,000 to more than 10,000 sources over the lifetime of the mission \citep{Nelemans2001b, Nelemans2001a, NelemansEtAl2004, LiuEtAl2010, RuiterEtAl2010}, providing the largest observational sample of these objects. At present it is not clear how the LISA sensitivity will be modified as a number of re-design options are being considered by the European Space Agency (ESA) and potential other partners. Regardless of the specifics it appears quite probable though that the sensitivity to Galactic DWDs at the middle to high-end frequency range relevant to this study will not be greatly affected.

Most of the studies in the literature concerning DWDs have been focused on circular binaries, which are expected to dominate the galactic DWD population. However, recent theoretical calculations by \citet{Willems2007} and \citet{Thompson2010} predicted a population of eccentric DWDs, which could provide a unique opportunity for investigating degenerate matter with LISA. In fact, the properties of WDs can be studied through the effects of tidal interactions on the detected GW signal from eccentric systems. \citet{Willems2007} studied eccentric binaries formed through dynamical interactions in globular clusters, (with eccentricities up to 0.8), while \citet{Thompson2010} studied eccentric DWDs as products of hierarchical triple systems, where the tertiary induces Kozai oscillations in the inner binary, driving it to high eccentricity (with eccentricities up to 1; see also \citealt{Gould2011} for a discussion of the implications of such a population).

In binaries with eccentric orbits, tidal forces cause a non-dissipative precession of the periastron of the orbit, also known as apsidal precession. This periodic motion 
leaves a mark in the GW signal that is potentially detectable. While in circular binaries GW radiation is emitted at multiples $n$ of the orbital frequency $\nu$, in the presence of periastron precession each of these harmonics is split into a triplet with frequencies $n\nu\pm\dot{\gamma}/\pi$ and $n\nu$, where $\dot{\gamma}$ is the apsidal precession rate \citep[hereafter WVK08]{Willems2008}. Various physical processes can lead to apsidal precession; in the specific case of tides it occurs because the tidal deformation of the components perturbs the stars' external gravitational field, and therefore the Keplerian motion of the binary components. 
Tidally-induced apsidal precession depends on the binary orbital parameters and on the internal structure of the tidally deformed stars. Therefore, comparisons between observed and theoretically predicted rates can be used to place constraints on the WD properties. An equivalent approach has been very successful in the study of binary main-sequence stars \citep{ClaretGimenez1993, ClaretGimenez2010, ClaretWillems2002, Schwarzschild1958}.  

Earlier studies carried out by WVK08 used zero-temperature WDs to investigate the contribution to apsidal precession rates due to general relativity (GR), tides, and rotation. WVK08 showed that, depending on the system parameters, tides could be the dominant contribution to apsidal precession in most of the LISA band.
The aim of this paper is to expand the initial study started by WVK08 and use detailed WD models from \citet{Deloye2007}, where the thermal properties of the WD are followed throughout the cooling phase, without zero-temperature assumptions. Even though we mainly focus our attention on detached DWDs, our results can be applied also to binaries containing a neutron star (NS) and a WD, which provide an even better probe to study WD physics.  In DWDs apsidal precession is due to both components, while in NS-WD binaries the distortion of the NS due to tides and rotation contributes negligibly, and apsidal precession carries the unique signature of the WD. We devote a section of the paper to a discussion of NS-WDs.

The plan of the paper is as follows. In \S\ref{sec:periPrecRatesTheo} we summarize the equations governing the tidal, rotational and GR contribution to apsidal precession. In \S\ref{sec:periPrecRatesCalc} we present the precession rates calculated using detailed WD models. We confirm the fundamental result previously obtained by WVK08 for zero-temperature models, and discuss the effect of WD temperature. In \S\ref{Sect:tidesPeriPrec} we study in more detail the tidally-induced apsidal precession rate. In \S\ref{Sect:SourcesMisclass} we discuss the implications of not properly accounting for tides when inferring WD masses from detected apsidal precession rates, while in \S\ref{Sect:CombiningObservations} we investigate the source parameters that could be constrained from a detection. In \S\ref{Sect:NSWD} we discuss apsidal precession in NS-WD binaries. We conclude in \S\ref{Sect:conclusions}.

\section{Apsidal Precession In Binaries} \label{sec:periPrecRatesTheo}
Here we summarize the equations governing the tidal, rotational, and GR contributions to apsidal precession in eccentric binaries. We consider a binary system consisting of two stars with masses $M_{1,2}$, and radii $R_{1,2}$, uniformly rotating with angular velocities $\Omega_{1,2}$. We take the axes of rotation to be orthogonal to the orbital plane. Let $\gamma$ be the argument of the periastron,  $P$ and $a$ the orbital period and semi-major axis, respectively, and $e$ the orbital eccentricity.
Following WVK08, we consider orbital and rotational periods long compared to the free harmonic periods of the component stars (the so-called static-tides regime; \citealt{Cowling1938, Sterne1939, SmeyersWillems2001}). In this regime we can neglect resonances between the tidal forcing angular frequencies and the eigenfrequencies of WD free oscillation modes.
Under these approximations, the contribution to the apsidal precession rate due to the quadrupole tides raised in star $i$ is \citep{Sterne1939}:
\begin{equation}
\dot\gamma_{Tid, i} = \frac{30\pi}{P}\left(\frac{R_i}{a}\right)^5\frac{M_{3-i}}{M_i}\frac{1+\frac{3}{2}e^2+\frac{1}{8}e^4}{(1-e^2)^5}k_i.
\label{eq:gammaDotTid}
\end{equation}
We assume that the WD components are sufficiently far apart that this quadrupole term dominates the precession (in \S\ref{subsect:gammaDotExample} we show that higher order terms indeed contribute negligibly).
As anticipated in \S\ref{Intro}, $\dot\gamma_{Tid}$ depends on the orbital parameters, component masses, and internal structure of the tidally deformed star. 
The dependence on the internal structure is embedded in the quadrupolar apsidal precession constant $k_i$, which measures the star's central concentration.
The constant $k_i$ ranges from 0 for a point mass to 0.74 for an equilibrium sphere with uniform mass density. For any realistic stellar model $k_i$ assumes values between 0 and 0.74.
In standard notation the quadrupole apsidal precession constant is normally referred to as $k_2$, where the subscript to $k$ refers to the dominant longitudinal mode in the spherical harmonic expansion of the tide-generating potential (as explained in Sect.\ref{sec:periPrecRatesCalc}). To avoid confusion, we will drop the subscript $2$, and use $k_i$ ($i$=1, 2) to denote the apsidal precession constant for each binary component.

A perturbation of the gravitational field and hence apsidal precession also arises because of the rotational distortion of the star caused by the centrifugal force. The rotationally-induced apsidal precession rate (e.g. \citealt{Sterne1939}) depends on the star's internal structure and radius in a similar fashion as $\dot\gamma_{Tid}$:
\begin{equation}
\dot\gamma_{Rot, i} = \frac{2\pi}{P}\left(\frac{R_i}{a}\right)^5\frac{M_1+M_2}{M_i}\frac{(\Omega_i/\Omega)^2}{(1-e^2)^2}k_i,
\label{eq:gammaDotRot}
\end{equation}
where $\Omega = 2\pi/P$ is the mean motion.

Finally, the GR contribution to apsidal precession at the leading quadrupole order is given by: 
\begin{equation}
\dot\gamma_{GR} = \left(\frac{2\pi}{P}\right)^{5/3}\frac{3G^{2/3}}{c^2}\frac{(M_1+M_2)^{2/3}}{(1-e^2)},
\label{eq:gammaDotGR}
\end{equation}
where $G$ and $c$ are the gravitational constant and the speed of light, respectively \citep{LeviCivita1937}. 
Note that $\dot\gamma_{GR}$ is independent of the internal structure of the components, and it is proportional to the total mass of the binary.

The total apsidal precession rate ($\dot{\gamma}$) is the sum of the three contributions above considering both WD components. 
If a detection of the apsidal precession rate in a GW signal were to be used to infer the source's total mass, it is evident that not accounting for tides and rotation could potentially lead to an overestimate of the system mass, and a possible misclassification of the GW sources (\S\ref{Sect:SourcesMisclass}).
\section{Apsidal Precession Rates} \label{sec:periPrecRatesCalc}
To investigate the various contributions to apsidal precession for eccentric DWDs in the LISA band, 
WVK08 considered zero-temperature WDs following a mass-radius relation as
 given by \citet{Nauenberg1972}, and setting the apsidal precession constant $k_i$ to 0.1 (appropriate for the WD masses considered in their study).
Here we use the detailed WD structure models from \citet{Deloye2007} to compute $k_i$ at different evolutionary phases during the WD lifetime.
\subsection{Apsidal Precession Constant $k_i$}
The apsidal precession constant is given by
\begin{equation}
2k_{i} = \frac{\xi_{i,T}(R_i)}{R_i}-1,
\end{equation}
where $\xi_{i,T}$ is the radial component of the tidal displacement of a mass element of star $i$.
In the limiting case of rotational and orbital periods long compared to the WD dynamical timescales (the so-called quasi-static tides regime), 
 $\xi_{i,T} (r)$ is a solution to the homogeneous second-order differential equation (e.g. \citealt{SmeyersWillems2001})
\begin{eqnarray} \nonumber
\frac{d^2\xi_{i,T}(r)}{dr^2}+2\left(\frac{1}{g(r)}\frac{dg(r)}{dr}+\frac{1}{r}\right) \\
\times\frac{d\xi_{i,T}(r)}{dr}-\frac{l(l+1)-2}{r^2}\xi_{i,T}(r) = 0,
\label{eq:Clairaut}
\end{eqnarray}
where $l$ is the longitudinal mode in spherical harmonics, and $g$ is the gravity. The quadrupole apsidal precession constant is obtained by setting $l=2$.
The desired solution to Eq. (\ref{eq:Clairaut}) should be finite at $r=0$ and at $r=R_i$ satisfy the boundary condition
\begin{equation}
\left(\frac{d\xi_{i,T}(r)}{dr}\right)_{R_i}+\frac{l-1}{R_i}\xi_{i,T}(R_i) = \epsilon_{i, T}(2l+1)c_{l, 0, 0}.
\label{eq:ClairautBC}
\end{equation}
Here $\epsilon_{i,T} = (R_i/a)^3(M_{3-i}/M_i)$ is a dimensionless parameter indicating the strength of the tidal force with
 respect to gravity at the star's equator, while $c_{l,0,0}$ are Fourier coefficients of degree $l$, depending on the
  star's radius as $(R_i/a)^{l-2}$.
The dominant contribution of the terms associated with the second-degree spherical harmonics for systems with components sufficiently far apart justifies 
restricting our analysis to the tides generated by the terms $l=2$.

\begin{figure} 
\plottwo{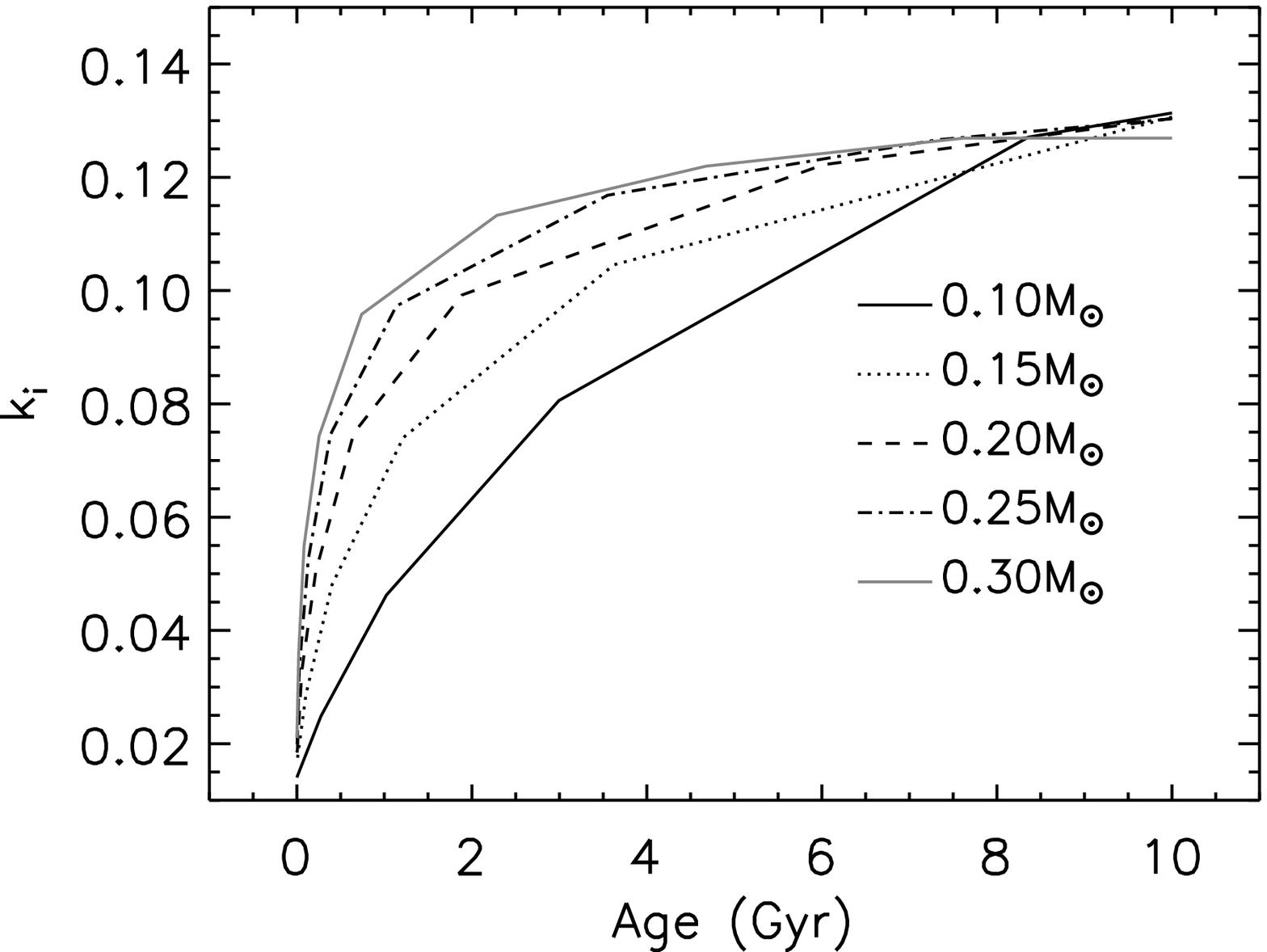}{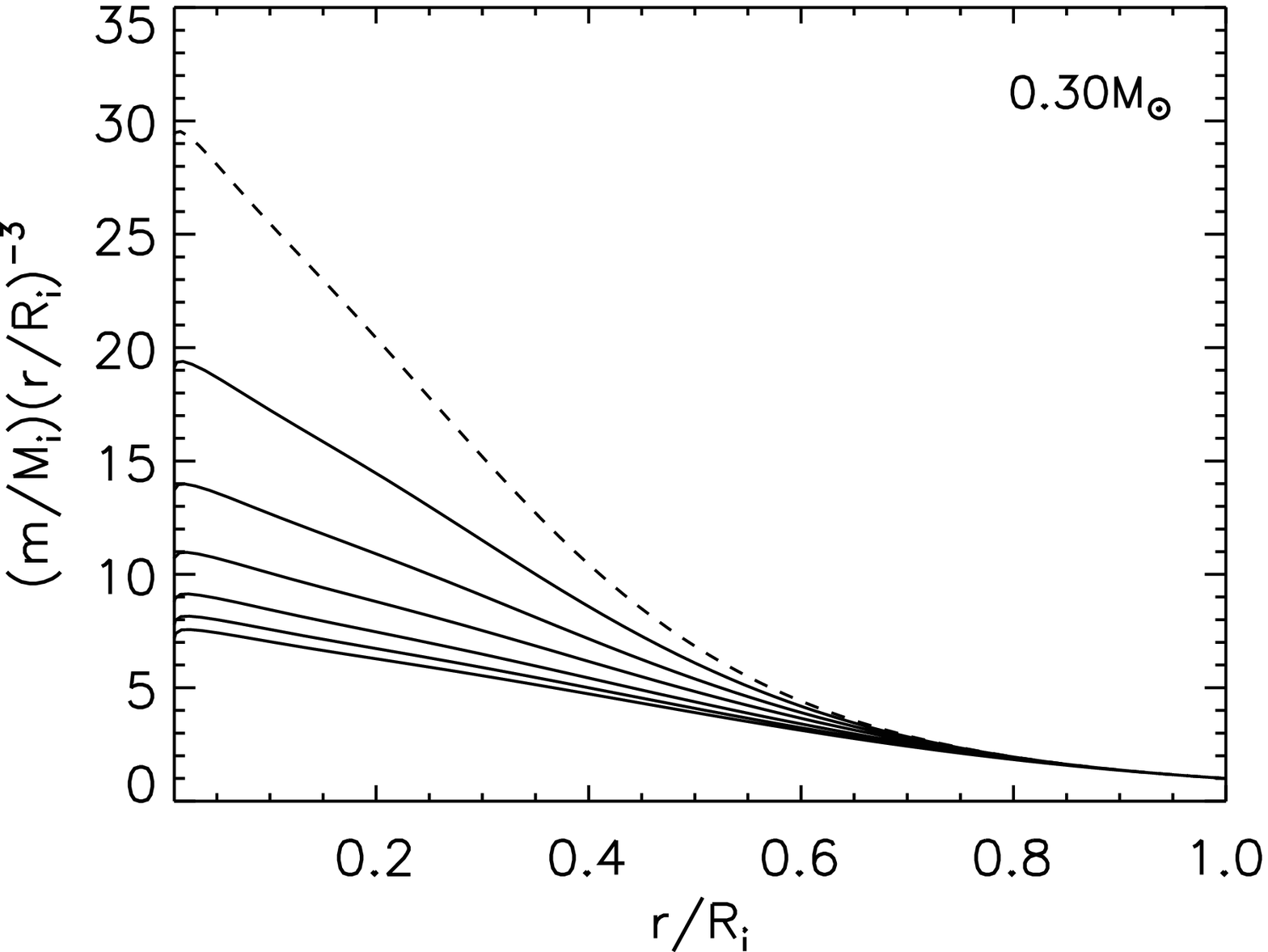}
   \caption{ \textit{Left: } Apsidal precession constants as a function of time for different He WD models. \textit{Right: } Radial density profile of a He WD with mass $M_i = 0.3\, M_\odot$ at different evolutionary stages. The dashed line represents the hot WD model and evolution in time proceeds downward. The flattening of the profile as the star cools is common to all WD masses. }
   \label{fig:k_density}
    \end{figure}

We consider He WD models from \citet{Deloye2007} with masses ranging from 0.1$\, M_\odot$ to 0.3$\, M_\odot$ in steps of 0.05$\, M_\odot$. For each mass we calculate the apsidal precession constant $k_i$ at different evolutionary stages. 
 The results are summarized in Table \ref{table:TeffAndk}, while in Figure \ref{fig:k_density} we show the time evolution of the apsidal precession
 constant for all the available masses (left panel). In what follows, we refer to the highest and lowest effective temperature ($T_{\rm{eff}}$) models as $\textit{hot}$, and $\textit{cool}$, respectively. For a given mass, a hot (cool) model represents a young (old) WD. 
We recall that $k_i$ measures the distribution of mass inside the star, and that it ranges from 0 for a point mass to 0.74 for a uniform density sphere. 
As shown in Figure \ref{fig:k_density}, $k_i$ increases with time, indicating that the WD becomes less centrally condensed while cooling. This trend is confirmed by the time evolution of the radial profile of the star's density, which becomes flatter as the WD evolves (right panel).
\begin{deluxetable}{cc|cc|cc|cc|cc}
\tablecolumns{5}
\tabletypesize{\scriptsize}
\tablewidth{0pc}
\tablecaption{Apsidal precession constants of WD models}
\tablehead{
 \multicolumn{2}{c}{0.1$\, M_\odot$}  & \multicolumn{2}{c}{0.15$\, M_\odot$} &  \multicolumn{2}{c}{0.2$\, M_\odot$}   & \multicolumn{2}{c}{0.25$\, M_\odot$}& \multicolumn{2}{c}{0.3$\, M_\odot$} \\
\colhead{ $T_{\rm{eff}} (K)$} & \colhead{$k$}   & \colhead{ $T_{\rm{eff}} (K)$}    & \colhead{$k$} &
\colhead{ $T_{\rm{eff}} (K)$} & \colhead{$k$}   & \colhead{ $T_{\rm{eff}} (K)$}    & \colhead{$k$}
& \colhead{ $T_{\rm{eff}} (K)$}    & \colhead{$k$}}
\startdata
5,448 & 0.014 & 9,743 & 0.017 & 14,903 & 0.017 &20,618 & 0.019 &26,527 & 0.021\\
5,327 & 0.025 & 9,106 & 0.029 & 13,233 & 0.033 &18,401 & 0.034 &23,598 & 0.036\\
4,436 & 0.046 & 7,475 & 0.048 & 10,874 & 0.050 &14,493 & 0.053 &18,279 & 0.055\\
3,203 & 0.081 & 5,340 & 0.074 & 7,735 & 0.075 &10,550 & 0.075 &13,524 & 0.074\\
2,335 & 0.127 & 3,520 & 0.105 & 5,166 & 0.099 &6,986 & 0.097 &8,995 & 0.096\\
1,899 & 0.147 & 2,463 & 0.134 & 3,271 & 0.122 &4,427 & 0.117 &5,747 & 0.113\\
1,640 & 0.153 & 2,174 & 0.142 & 2,639 & 0.134 &3,351 & 0.127 &4,323 & 0.122\\
	        &	     & 1,989 & 0.146 & 2,422 & 0.139 & 2,871 & 0.133 & 3,585 & 0.127\\
\enddata
\label{table:TeffAndk}
\end{deluxetable}
\subsection{An Example of Apsidal Precession Rates}\label{subsect:gammaDotExample}
\begin{figure} 
\plotone{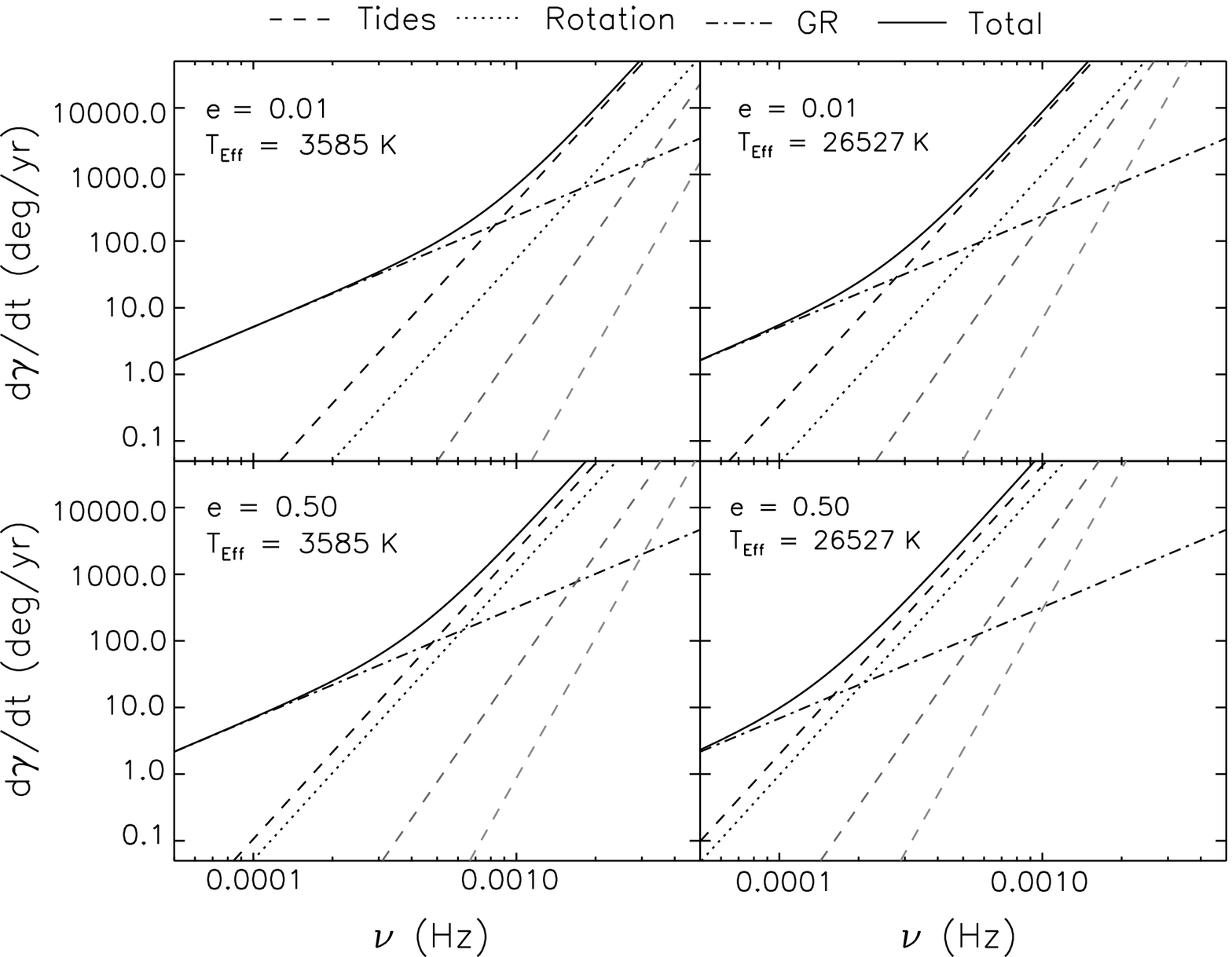}
   \caption{Apsidal precession rates for cool (\textit{left}) and hot (\textit{right}) DWDs with masses $M_1 = M_2 = 0.3\, M_\odot$, and orbital eccentricity $e = $0.01 (\textit{top}), and $e = $0.5 (\textit{bottom}). The tidal and rotational rates include the contribution of both stars. The dark- and light-grey dashed lines represent the tidal contribution from the longitudinal modes $l=3$ and $l=4$, respectively. Given the dependency of $\dot\gamma_{Tid, i}$ and $\dot\gamma_{Rot, i}$ on $R_i^5$, the contraction of the stars with time is responsible for the decrease of these two contributions with the evolutionary stage of the WDs. In the frequency range considered, the cool DWD at $e = $0.5 begins Roche-lobe overflow (RLO) at $\nu \simeq 2.7\,$mHz (at $e = 0.01$ the onset of RLO is at a higher frequency than displayed), while the hot DWD at $e = $0.01 (0.5) begins RLO at $\nu \simeq 1.8 (0.65)\,$mHz.}
   \label{fig:periPrec}
    \end{figure}
In Figure \ref{fig:periPrec} we show an example of tidal, rotational, and GR apsidal precession rates calculated with our models, as a function of the orbital frequency
 $\nu = 1/P$ for different orbital eccentricities and WD masses of 0.3$\, M_\odot$. 
The rotational contributions are calculated assuming pseudo-synchronism, i.e. that each WD's rotational angular velocity is equal to the orbital angular velocity at periastron (a note on this assumption follows below). The masses were chosen in order to compare our calculated apsidal precession rates against WVK08's results for zero-temperature WDs, with which our conclusions in general agree.  We note that apsidal precession is dominated by GR for systems with small orbital frequency, while tides and rotation become the dominant contributions towards higher frequencies. As a reference, for the tidal contribution we include also the rates associated with the $l=3$ and $l=4$ longitudinal modes \citep{Sterne1939}. We note that the total phase shift to the signal from these terms is about a factor of ~100 to ~1,000 times smaller than the $l = 2$ tidal
contribution. If synchronization at periastron is assumed, tides always dominate rotation. The same is true for a sub-synchronous system, given the dependency of Eq. (\ref{eq:gammaDotRot}) on the WD's rotational angular frequency. 

Observational constraints derived both from traditional spectroscopy and asteroseismology add to the growing evidence that single WDs are slow rotators \citep{Heber1997,  Koester1999, Berger2005, Kawaler2004}. The only DWD system for which there has been an estimate of the rotational velocity of the components  is SDSS J125733.63+542850.5. For this binary, \citet{Marsh2010} and \citet{Kulkarni2010} determined the rotation 
rate of both WDs to be super-synchronous, assuming the observed broadening of the spectral lines were solely due to rotation. However both studies cannot exclude that the broadening seen could be due, instead, to strong 
magnetic fields in more slowly rotating stars (\citealt{Kulkarni2010} and Marsh 2010, private communication). Further observations are needed in order to settle this issue.
Lacking more elaborate evidence that super-synchronous rotation is the norm, in this work we assume that synchronization at periastron is an upper limit on DWD rotation.

It is clear from Figure \ref{fig:periPrec} that apsidal precession is significant in most of the LISA band for both hot and cool DWDs, and high
  and low orbital eccentricities. At $\nu \simeq 1\,$mHz, 
 $e = 0.01$, and for cool DWDs  the precession of the periastron is expected to shift the phase of a GW signal by 3$\pi$-4$\pi$ over a 1$\,$yr mission, and is therefore detectable (WVK08). The shift increases by more than a factor of 10 for hot WDs, and, at the same frequency of $\simeq 1\,$mHz, tides are the leading mechanism (accounting for $\sim 85\%$ of the total $\dot{\gamma}$).
For higher orbital eccentricities, tides dominate $\dot\gamma$ at shorter orbital frequencies leading to much higher rates.
We also note that, given the dependence of $\dot\gamma_{Tid, i}$ on $R_i^5$ binaries hosting hot (and bigger) WDs have $\dot{\gamma_{Tid}}$ systematically higher than cool (and contracted) DWDs by a factor of $\sim 20$:  at $\nu \simeq 1\,$mHz and $e = 0.01$, $\dot\gamma_{Tid, i}$ for hot(cool) DWDs is $\simeq7300(390)\,$deg/yr.

We note that among the available models, a DWD hosting 0.3$\, M_{\odot}$ components yields the lowest $\dot\gamma_{Tid, i}$ and $\dot\gamma_{Rot, i}$, and the highest $\dot\gamma_{GR, i}$. In fact, the larger radii of the lower mass models act to increase the tidal and rotational contributions relative to GR. The relative GR contribution, instead, being dependent upon the total system mass, decreases with the components' masses.
\section{Tidal Contributions to Apsidal Precession}\label{Sect:tidesPeriPrec}
GR-induced apsidal precession, if detected in binaries with long orbital periods (where this contribution is dominant), can be used to constrain the total system mass (see Eq. \ref{eq:gammaDotGR}). Here we show that $\dot\gamma_{Tid, i}$ can be used in a similar fashion.

\begin{figure} 
\plotone{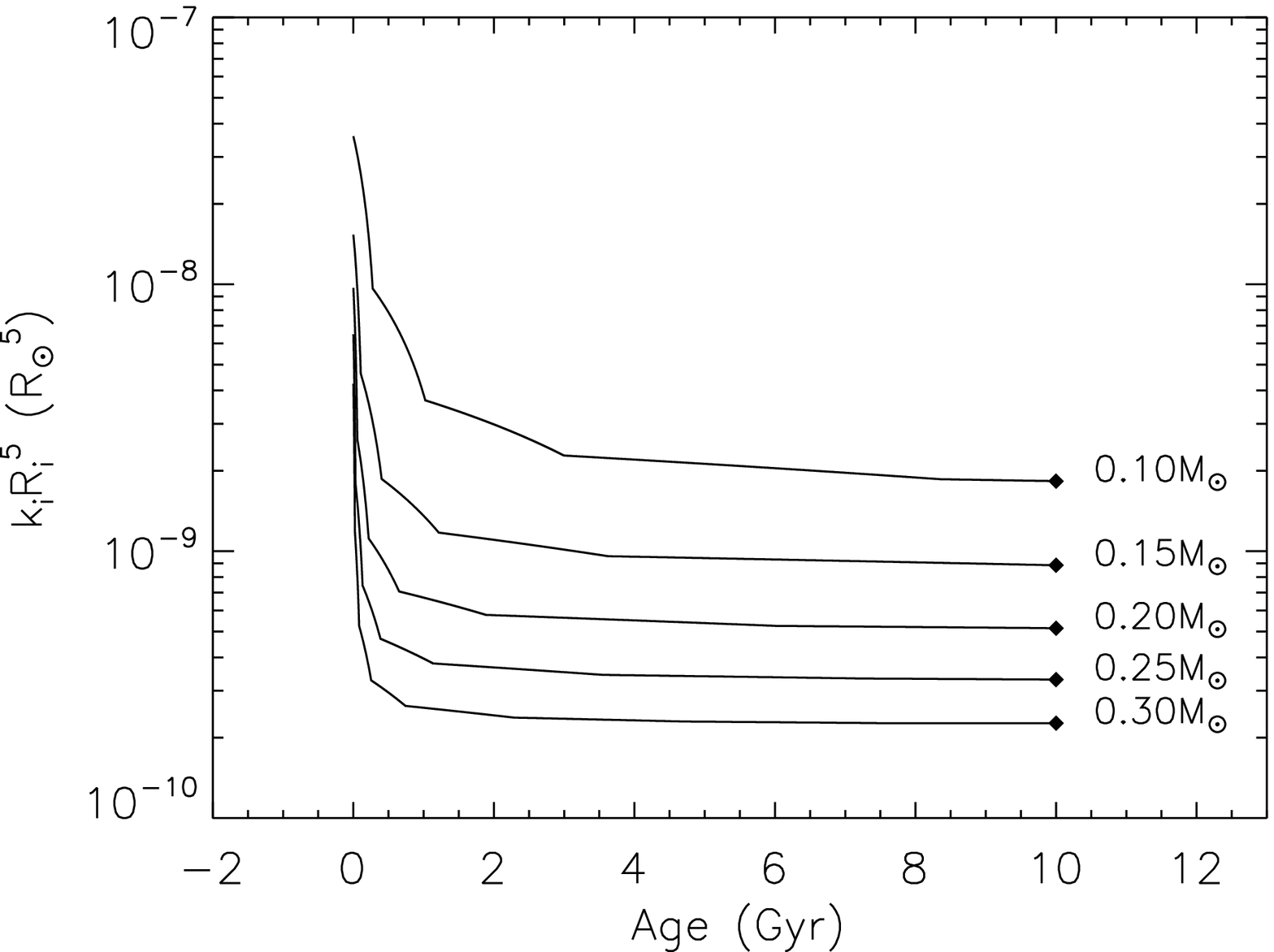}
   \caption{Evolution of $k_iR_i^5$ as a function of the WD's age. Filled squares denote plateau values. The maximum age considered for each model is 10$\,$Gyr.}
   \label{fig:k2R5_age}
    \end{figure}

\begin{figure} 
\plotone{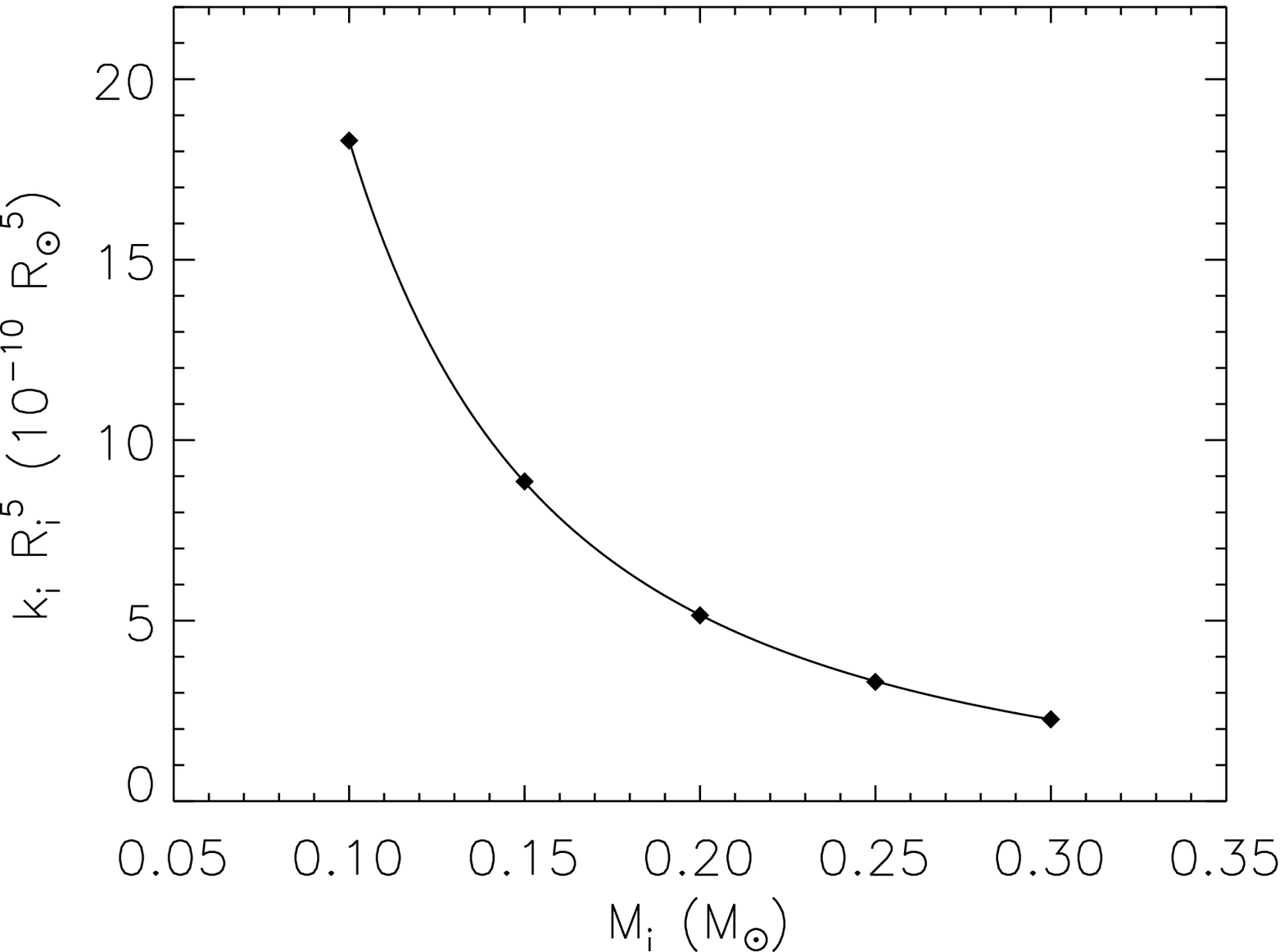}
   \caption{Calculated $k_iR_i^5$ as a function of the WD mass. Filled squares represent the plateau values of $k_iR_i^5$. The solid line is the fit given by Eq. (\ref{eq:k2r5_min_fit}).}
   \label{fig:k2R5_mass}
    \end{figure}

Given the dependence of Eq. (\ref{eq:gammaDotTid}) on several stellar and binary properties ($P, e, a, M_1, M_2$, and $R_i$), an apsidal precession detection and rate measurement cannot be used to place tight constraints on the binary properties. 
In fact, although the orbital period and eccentricity can be precisely determined from the frequency spectrum of the GW signal, measuring radii and masses is much more challenging.  
In what follows, we show that some of the unknowns determining $\dot\gamma_{Tid, i}$ can be eliminated if one considers the time evolution of the term $k_i R_i^5$ that enters Eq. (\ref{eq:gammaDotTid}). As Figure \ref{fig:k2R5_age}  shows, during most of the WD's lifetime $k_i R_i^5$ assumes a plateau value which is unique for each mass. Consequently Eq. (\ref{eq:gammaDotTid}) effectively depends only on the orbital period, eccentricity, and component masses (through Kepler's law $a$ can be re-written as a function of $P$, $M_1$,  and $M_2$).
According to our models, the plateau values of $k_iR_i^5$ (filled squares in Figure \ref{fig:k2R5_age}) depend on the WD mass as follows:
 \begin{equation}
k_iR_i^5 = -0.632+0.370\cdot M_i^{ -1.709},
\label{eq:k2r5_min_fit}
\end{equation}
where $M_i$ is the mass in solar units, and $k_iR_i^5$ is in $10^{-10} R_\odot^5$ (see Figure \ref{fig:k2R5_mass}). 
    
\begin{figure} 
\plotone{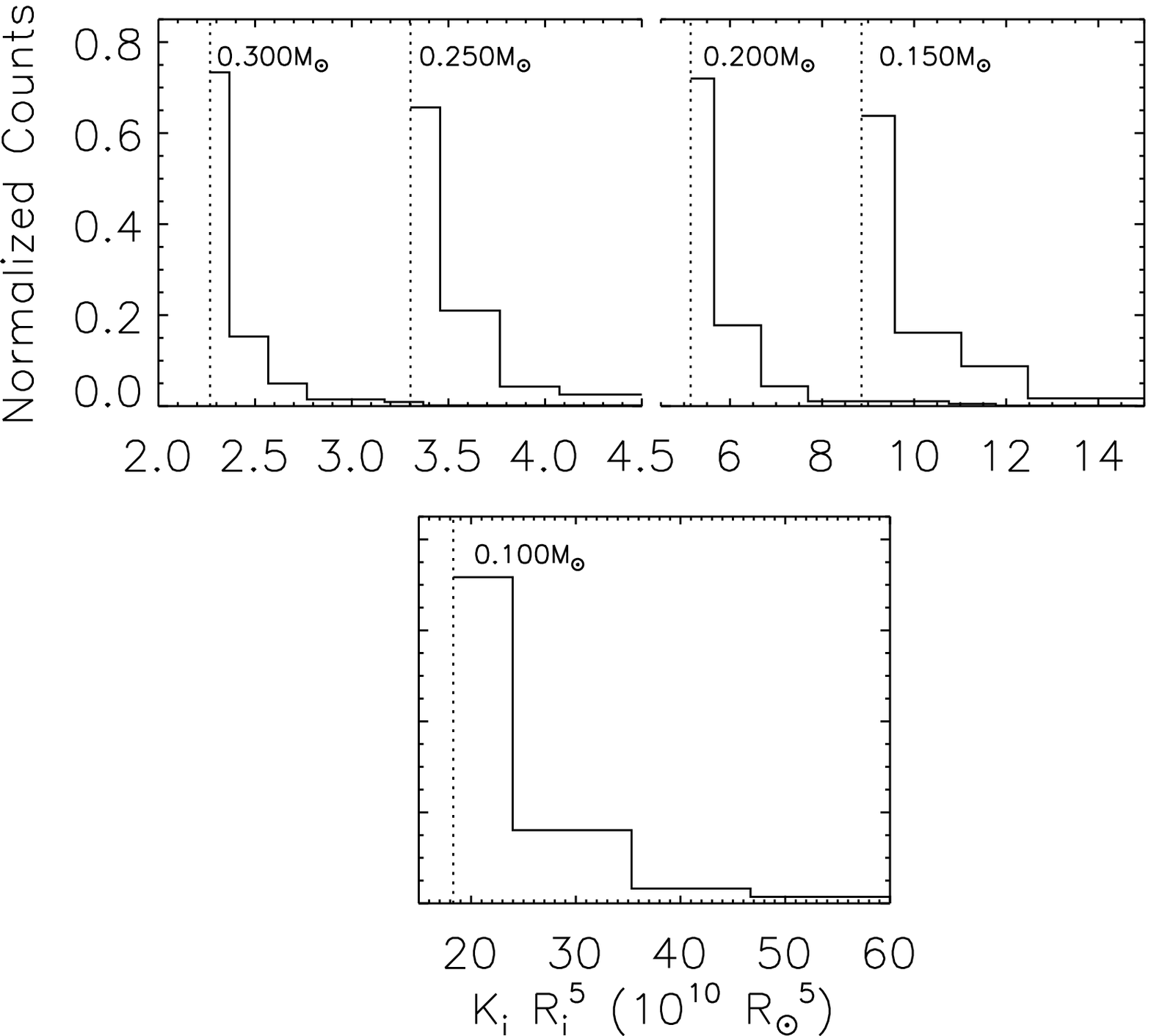}
   \caption{PDF for $k_iR_i^5$ for each model mass assuming an observational time uniformly distributed between 0 and 10$\,$Gyr after WD formation. The vertical dotted line represents the plateau of $k_iR_i^5$ for each mass. Note the non-uniform x-axis.}
   \label{fig:PDF_k2R5}
    \end{figure}

Even though for most of the WD lifetime there is a unique relation between $k_iR_i^5$ and $M_i$ that we can exploit to simplify Eq. (\ref{eq:gammaDotTid}), we note that the uniqueness of $k_i R_i^5$ for a given mass is broken for young WDs. The probability density functions (PDFs)  for the values of $k_iR^5$ assuming an observational time uniformly distributed between 0 and 10$\,$Gyr after WD formation (consistent with our current understanding of star formation rate in the Milky Way) are shown in Figure \ref{fig:PDF_k2R5}. The dotted 
 vertical lines represent the plateau of $k_iR_i^5$ for each mass, which is used for the one-to-one correspondence to the WD mass. In Table \ref{table:PDFk2r5} we summarize the percent of overlap for the calculated PDFs of $k_iR_i^5$, which represents the probability of confusing a less massive old (cool) WD with a more massive young (hot) WD.
We conclude that, for a given $k_i R_i^5$, we can distinguish between two masses within 0.05$\, M_\odot$ at the 90-95\% confidence level, without any information on the age of the WDs. 

Given the unique dependence of $k_iR_i^5$ on the mass for cool WDs, the tidal contribution to apsidal precession can now be used to place constraints on the component masses also for short period binaries, extending this possibility to the entire LISA frequency domain. The substitution of Eq. (\ref{eq:k2r5_min_fit}) in Eq. (\ref{eq:gammaDotTid}) makes $\dot\gamma_{Tid, i}$ depend on only $P$, $e$, and $M_i$, and since $P$ and $e$ can
be precisely determined from the GW signal a measurement of $\dot\gamma$ yields constraints on a particular combination of the masses in the
system, which we call the "apsidal mass function." 
\begin{deluxetable}{ccc}
\tablecolumns{5}
\tabletypesize{\scriptsize}
\tablewidth{0pc}
\tablecaption{Probability of overlap in the values of $k_iR_i^5$ between an old, lower-mass WD with a young higher-mass WD, assuming
observations uniformly distributed in the WD age.  This probability is calculated from the time evolution of $k_i R_i^5$ appearing in Figure \ref{fig:k2R5_age}.}
\tablehead{
\colhead{Higher Mass ($M_\odot$)}  & \colhead{Lower Mass ($M_\odot$)}    & \colhead{ Confusion Probability (\%)}}
\startdata
0.3 & 0.25& 2.51 \\
0.3 & 0.2 & 0.91 \\
0.3 & 0.15 & 0.49 \\
0.3 & 0.1& 0.16  \\
0.25 & 0.2& 3.44 \\
0.25 & 0.15& 1.17\\
0.25 & 0.1& 0.30   \\
0.2 & 0.15& 4.61  \\
0.2 & 0.1& 1.41\\
0.15 & 0.1& 4.43\\
\enddata
\label{table:PDFk2r5}
\end{deluxetable}

\section{Misclassification of the Sources}\label{Sect:SourcesMisclass}

\begin{figure} 
\plottwo{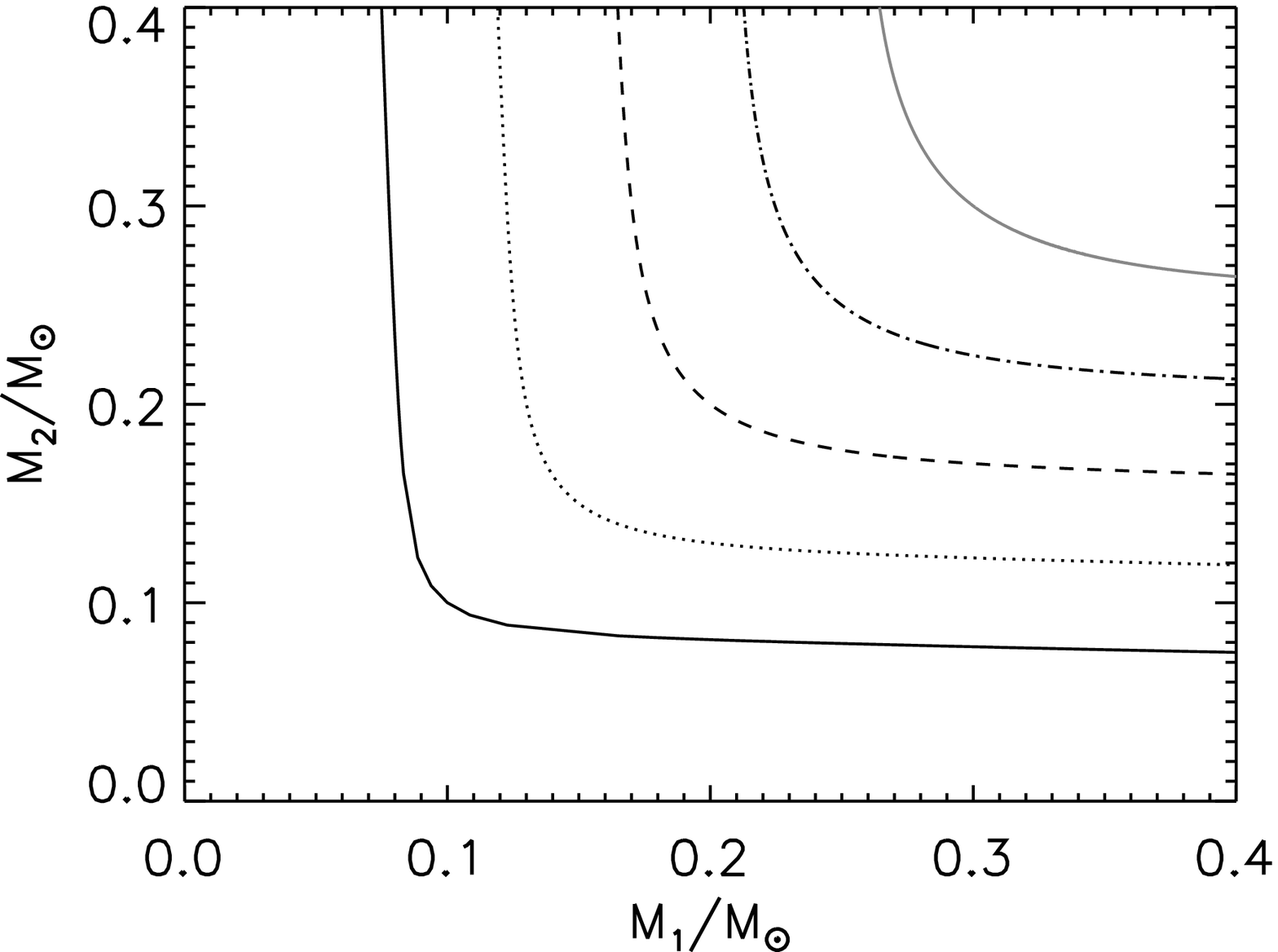}{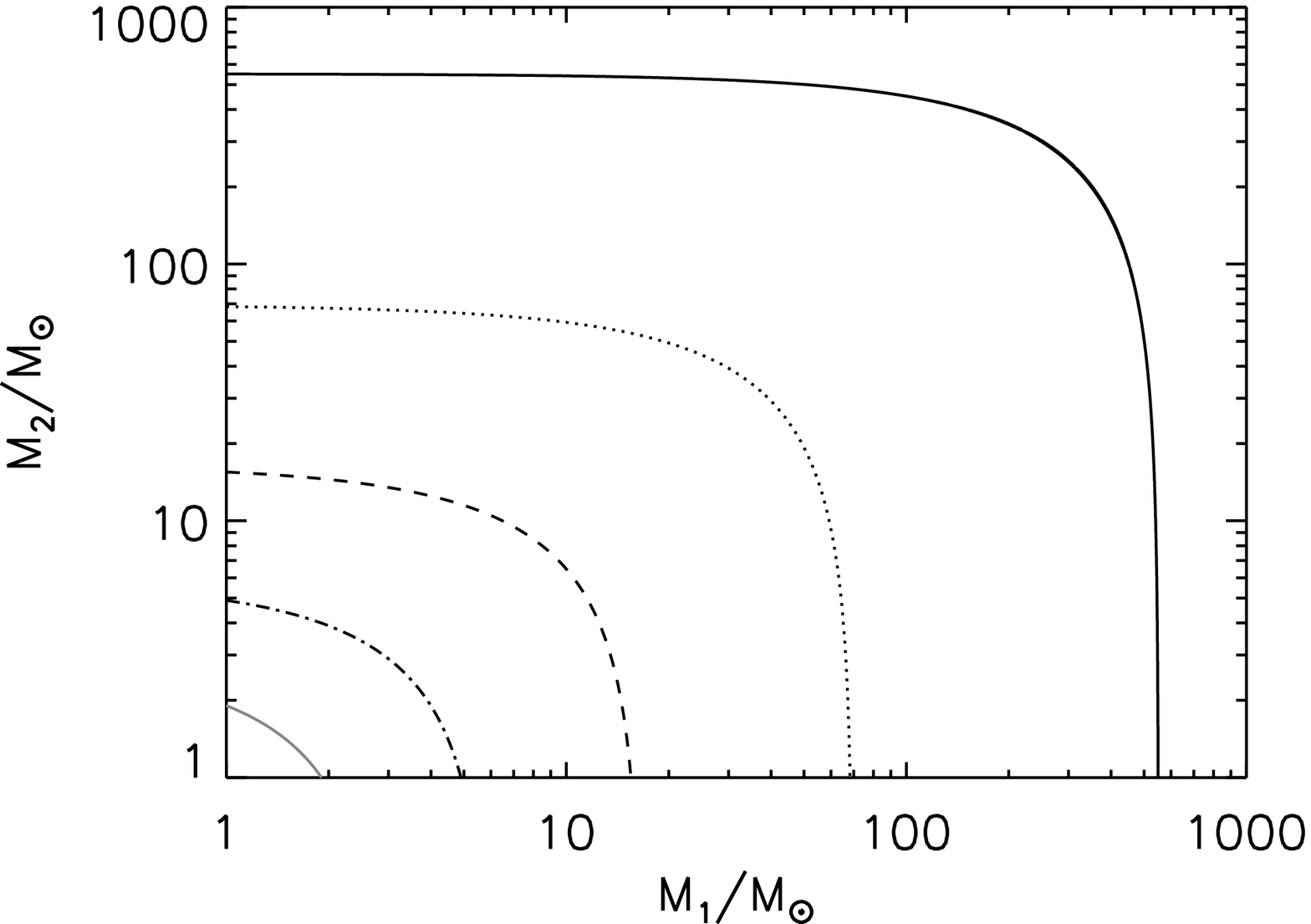}
   \caption{Lines of constant apsidal precession rate calculated for cool DWDs with orbital frequency $\nu=1\, mHz$, and eccentricity $e = 0.01$. \textit{Left: } $\dot\gamma$ calculated for different combinations of components' masses ($M_1$ and $M_2$) accounting for GR, rotation and tides. \textit{Right: } components' masses that would be inferred if only GR was taken into account. The various lines correspond to $\dot\gamma$ in deg/yr of $\simeq$ 22570 (black-solid), $\simeq$ 5700 (dotted), $\simeq$ 2200 (dashed), $\simeq$ 1100 (dot-dashed), $\simeq$ 680 (grey-solid).}
   \label{fig:DetectionError}
    \end{figure}

Ignoring tidal effects when trying to constrain the apsidal mass function from a measured rate of apsidal precession would lead to a dramatic overestimate in the masses, as shown in Figure \ref{fig:DetectionError}. The left panel depicts lines of constant total $\dot{\gamma}$ (i.e. GR, tides, and rotation) for cool DWDs with orbital frequency $\nu=1\, mHz$, and eccentricity $e = 0.01$. The tidal and rotational rates include the contribution of both stars. The right plot shows, for these same apsidal precession rates, the masses that would be inferred solely accounting for GR. Unless the distance to the source is known (see \S\ref{Sect:CombiningObservations}), the amplitude of the signal gives no constraint on the masses, which would therefore be overestimated by orders of magnitude when only accounting for the GR parts of the apsidal mass function.  In this case,
an eccentric WD binary would be misclassified as a binary BH or NS. Such a misclassification would bias the ratios of different populations of compact object binaries that hold essential information about binary formation mechanisms and stellar evolution, and would hamper a correct astrophysical interpretation of the source's nature and properties.

\section{Adding $\dot{\gamma}$ to GW Data Analysis: Constraining the Source Properties}\label{Sect:CombiningObservations}

The GW signal emitted by DWDs at GW frequencies $f  (= 2\nu)\approx \,$ a few mHz is expected to show an intrinsic frequency evolution \citep{WebbinkHan1998,Willems2007}. We now discuss how a detection of apsidal precession in such binaries can be used to constrain the individual masses of the WD components.

The measurable signal for GW detectors is the strain amplitude of the wave:
\begin{eqnarray} \nonumber
h(n, e) = 1.0\times 10^{-21}\frac{\sqrt{g(n, e)}}{n}\\
\times\left(\frac{\mathcal{M}}{M_\odot}\right)^{5/3}\left(\frac{P}{1hr}\right)^{-2/3}\left(\frac{d}{1kpc}\right)^{-1}.
\label{eq:strainAmplitude}
\end{eqnarray}
The pre-factor $\sqrt{g(n, e)}/n$ gives the scale of the strain amplitude produced by the $n$th orbital harmonic at eccentricity $e$ (e.g., \citealt{NelemansEtAl2001}), $P$ is the orbital period, $d$ is the distance to the source, and $\mathcal{M} = (M_1M_2)^{3/5}/(M_1+M_2)^{1/5}$ is the so-called chirp mass.
If radiation reaction causes the orbit to evolve (and therefore the intrinsic frequency of radiation to change), assuming this drift is GR-driven, and that tidal and/or magnetic spin-orbit coupling do not contribute, the chirp mass can be determined from (e.g. \citealt{Schutz1996})
\begin{equation}
\dot{f} = 5.8\times10^{-7} \left(\frac{\mathcal{M}}{M_\odot}\right)^{5/3}f^{11/3} Hz\, s^{-1}.
\label{eq:FreqEvolution}        
\end{equation}
The distance to the source can then be calculated via Eq. (\ref{eq:strainAmplitude}).
It is clear that an apsidal precession detection combined with a determination of the chirp mass yields the masses of the individual components: $\mathcal{M}$ and Eq.  (\ref{eq:gammaDotTid}) or (\ref{eq:gammaDotGR}) for tidally- or GR-induced apsidal precession, and $\mathcal{M}$ together with Eq. (\ref{eq:gammaDotTid})+(\ref{eq:gammaDotGR}) in the regime where tides and GR contribute equally are sufficient to determine the
individual masses of the components. Given the currently observed sample of rotation rates in isolated WDs and DWDs, we can assume that GR and tides are the only mechanisms contributing to apsidal precession.

In the event that a signal has measurable $\dot{f}$, the problem of source misclassification discussed in \S\ref{Sect:SourcesMisclass} is less acute.  The chirp mass implied by the measured $\dot{f}$ would be inconsistent with the large system mass inferred from the GR contribution to apsidal precession alone; this inconsistency can be resolved, and the proper component masses determined, by properly including the tidal contribution to apsidal precession as discussed above.
\section{Neutron Stars - White Dwarfs Binaries}\label{Sect:NSWD}
More convenient laboratories for investigating WD physics via apsidal motion are NS-WD binaries. As opposed to DWDs, where the tidal and rotational contribution of both components have to be taken into account, apsidal precession in NS-WDs carries solely the signature of the WD, because the tidal and rotational distortion of the NS do not contribute to the apsidal precession.
Eccentric NS-WDs are observed in the field typically as radio pulsars orbiting WDs \citep{Lorimer2005}, and two such binaries have been discovered to date (PSR B2303+46, \citealt{StokesEtAl1985, KerkwijkKulkarni1999}, and PSR J1141Ð6545, \citealt{KaspiEtAl2000, Bailes2003}).
This field population of eccentric NS-WDs is thought to have formed from the evolution of binaries with similar mass components. The originally more massive primary transferred mass to its companion to become a WD, while the secondary accreted enough matter to evolve into a NS, kicked at supernova into an eccentric orbit \citep{TutukovYungelSon1993}. Even though no such systems have been observed in the LISA frequency band, theoretical simulations predict a wide range of orbital periods and eccentricities, and a subset of this population might be readily detectable by LISA (\citealt{KalogeraEtAl2005} and references therein).

Another advantage of using NS-WDs to investigate WD physics via apsidal precession lies in the higher number of expected systems if compared with eccentric DWDs in globular clusters. In fact, the expected formation rate of galactic eccentric NS-WDs is between 10 to 10,000 times the expected formation rate of eccentric DWDs in globular clusters (see \citealt{KalogeraEtAl2005} and references therein for a comparison between empirical and theoretical rate estimates, \citealt{Willems2007}).

\begin{figure} 
\plotone{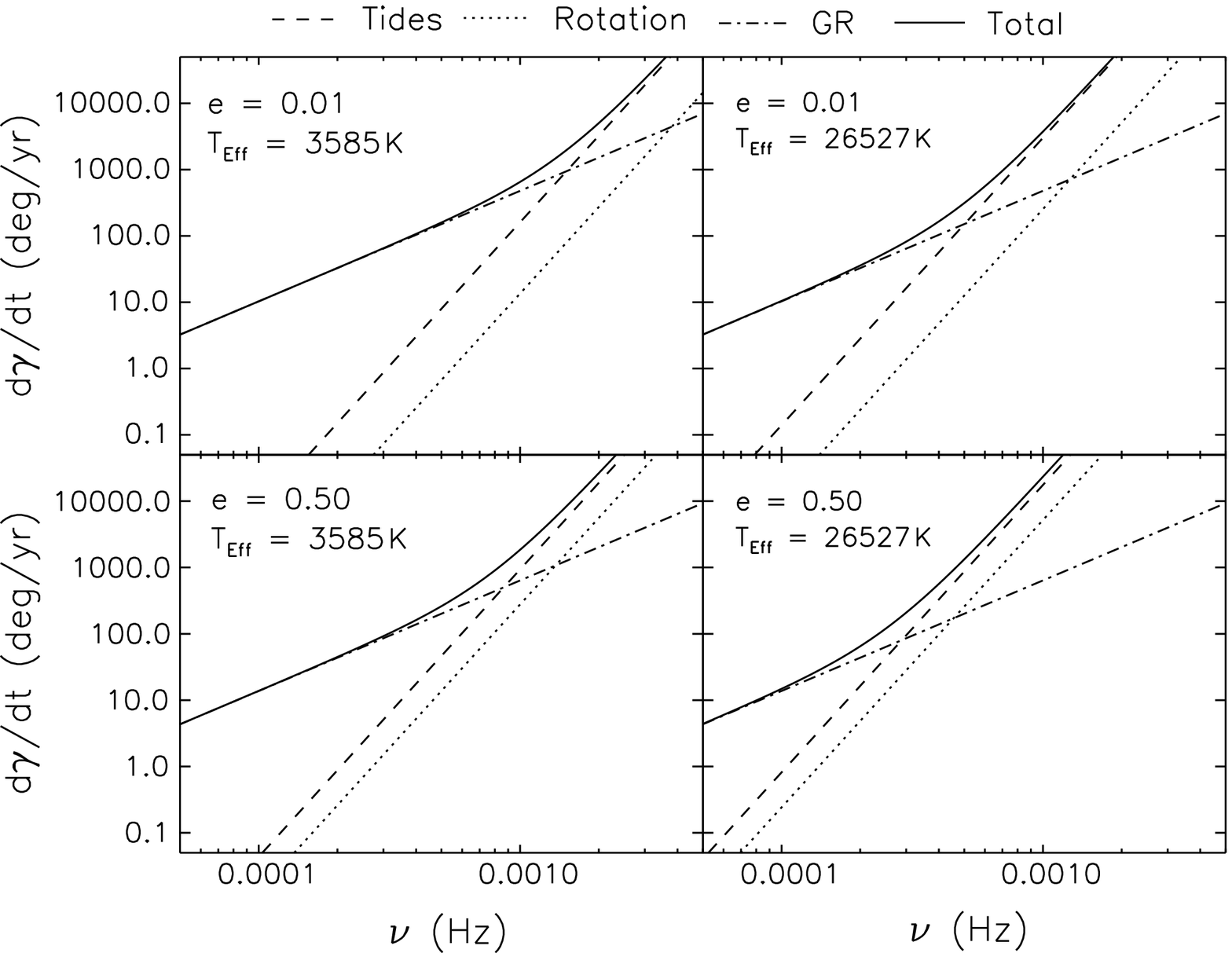}
   \caption{Same as Figure \ref{fig:periPrec}, but for a NS-WD binary hosting a 1.4$\, M_\odot$ NS and a 0.3$\, M_\odot$ WD. In the frequency range considered, the cool WD at $e = $0.5 begins RLO at $\nu \simeq 2.5\,$mHz (at $e = 0.01$ the onset of RLO for a cool WD is at a higher frequency than displayed), while the hot WD at $e = $0.01 (0.5) begins RLO at $\nu \simeq 1.7 (0.6)\,$mHz.}
\label{fig:periPrecNSWD}
\end{figure}
Figure \ref{fig:periPrecNSWD} shows, as an example, the tidal, rotational, and GR apsidal precession rates as a function of the orbital frequency for NS-WD binaries with different orbital eccentricities. To make a comparison with Figure \ref{fig:periPrec} we consider a 0.3$\, M_\odot$ WD ($M_1$), while assuming a typical mass of 1.4$\, M_\odot$ for the NS ($M_2$). The rotational contribution of the WD is calculated assuming synchronization at periastron.  
Again, GR dominates apsidal precession in binaries with the longest orbital period, while tides become progressively more important as the frequency increases. When compared to the DWD binaries shown in Figure \ref{fig:periPrec}, the dependency of Eqs. (\ref{eq:gammaDotTid}), (\ref{eq:gammaDotRot}), and (\ref{eq:gammaDotGR}) on $M_2$ leads lower tidally- and rotationally-induced apsidal motions, and a higher GR contribution\footnote{If $M_1, P$, and $e$ are kept fixed, using Kepler's law $\dot{\gamma}_{Tid,i}\propto M_2(M_1+M_2)^{-5/3}$, $\dot{\gamma}_{Rot,i}\propto (M_1+M_2)^{-2/3}$, and $\dot{\gamma}_{GR,i}\propto(M_1+M_2)^{2/3}$}. Consequently, for the same orbital configuration and WD masses, the predominance of tides in a NS-WD occurs at higher orbital frequencies than in a DWD. Nevertheless, the tidal contribution needs to be taken into account to properly interpret the observations.

A more detailed analysis of apsidal precession in eccentric NS-WD binaries will be the subject of future work. However, here we note that these systems potentially constitute an important astrophysical laboratory for testing our models. 
In NS-WDs, pulsar timing measurements might yield the masses of the individual components independently of precession effects, and therefore allow a comparison between the apsidal precession prediction of models and measurements.
\section{Summary, Discussion and Conclusions}\label{Sect:conclusions}
In this paper we use detailed WD models to continue the investigation started by WVK08 on apsidal precession due to tides, rotation and GR in eccentric DWDs.
We find that apsidal motion can lead to a significant shift in the emitted GW signal, the effect being stronger for binaries with hot WDs, and weaker for cool WDs.

In general, GR is the dominant mechanism at small orbital frequencies, while tides and rotation become increasingly important as the frequency increases. 
Given our current understanding of WD rotation rates, we can safely assume that apsidal precession is solely tidally-induced at frequencies where GR doesn't dominate the precession. 
Our detailed models also show that $\dot\gamma_{Tid, i}$ in binaries hosting young WDs is systematically higher than in binaries hosting old WDs by a factor of $\sim 20$, but note that WDs spend only a small fraction of their lifetime at these temperatures.

We investigate the astrophysical information that can be extracted from measured apsidal precession rates. 
First, we find that the component masses that would be inferred from the GR contribution alone could be overestimated by orders of magnitude. 

Secondly, we show that, as the WD ages, $k_iR_i^5$ asymptote to a value that depends only on the WD mass. This behavior allows us to simplify the form of Eq. (\ref{eq:gammaDotTid}), and to use  $\dot\gamma_{Tid, i}$ to place constraints on the apsidal mass function for short-period binaries. As noted by WVK08, the leading GR contribution in long-period DWDs yields the total system mass, therefore our investigation extends the possibility of measuring the apsidal mass function to the entire LISA frequency domain.
Furthermore, for systems where the distance can be independently measured via electromagnetic observations, or in which radiation reaction causes the orbit to evolve, apsidal precession and chirp mass measurements can be used to determine the masses of the individual components. 

Apsidal precession in eccentric NS-WD systems carries more straightforward information about the WD mass. These systems are formed in the field at rates 10-10,000 higher than that of eccentric DWDs in globular clusters. In these systems, the tidal and rotational distortion of the NS component contribute negligibly to $\dot{\gamma}$, and apsidal motion carries the unique signature of the WD. We show that in NS-WD binaries apsidal precession can also lead to a considerable shift in the GW signal in most of the LISA band. The  electromagnetic detection of these binaries as a WD orbiting a radio pulsar, combined with a possible determination of the components masses via timing measurements, provide a testbed for validating the reliability of our WD models.

In this paper we investigate the importance of  tidally-driven apsidal precession in low-frequency GW sources. However, tidal effects are also being discussed in the literature for high-frequency sources ($f \gtrsim 10\,Hz$), such as NS-NS binaries (DNS). DNS are expected sources for the next generation of ground-based GW detectors (e.g. Advanced LIGO; \citealt{Harry2010}). In such binaries, the tidal distortion of the NS components and its imprint on the emitted GW signal during the inspiral phase could potentially be used to extract information about the NS equation of state (EOS), and therefore the NS matter (e.g. \citealt{Flanagan2008, Hinderer2010}). Here we note that in the static-tides regime the leading mechanism to apsidal precession in DNS is GR, and that the tidal and rotational contributions to $\dot{\gamma}$ can not be used to place constraints on the stars' central concentration. To give a sense of the magnitude of the various contributions to $\dot{\gamma}$ in DNS, we consider a binary hosting two 1.4$\,M_{\odot}$ NS emitting GWs at a frequency of 10$\,$Hz ($P = 200\, ms$). To maximize the tidal and rotational contributions to apsidal precession, we consider the biggest NS radius given by one of the commonly used EOS ($\simeq14\,$Km).  \cite{Hinderer2010} estimated the apsidal precession constant to be $k_2\simeq $0.1. For this configuration of masses and radii, we consider the highest eccentricity not to have Roche-lobe overflow at periastron according to \citet{SepinskyEtAlNoFred2007}, $e\simeq $0.85. Furthermore, to maximize the rotational contribution (recall that $\dot\gamma_{Rot, i}\propto (\Omega_i/\Omega)^2$), we consider a rotational period of 20$\,$ms for both components (\citealt{Reisenegger1992} predicted the g-modes period in NSs to be as low as 10$\,$ ms). In this configuration, $\dot{\gamma_{GR}}$ is $\sim 10^3$$\dot{\gamma_{Tid}}$ and $\sim 10^4$$ \dot{\gamma_{Rot}}$. Obviously, the tidal contribution increases at the orbit shrinks. However, also decreasing the orbital period to 50$\,$ms, while keeping the rotational period the same, $\dot{\gamma_{GR}}$ remains two orders of magnitude bigger than the tidal contribution.

We conclude that the inclusion of tides is vital to the proper analysis of detected apsidal motion rates of DWDs and NS-WD binaries in the LISA band. DWDs, the most numerous guaranteed sources for LISA, hold essential signatures of binary and stellar evolution mechanisms. The possibility of detecting apsidal precession in these binaries opens up a new and exciting window on WD physics that could not be investigated otherwise.
 
\begin{acknowledgements}
We thank M. Benacquista, J. R. Gair, I. Mandel, and A. Sesana for useful discussions, and the anonymous referee for his/her positive and constructive review.
This work was supported by NASA Award NNX09AJ56G.
\end{acknowledgements}
\newpage
\bibliographystyle{apj}

\end{document}